\documentclass[twocolumn,aps,prl]{revtex4}
\usepackage{graphicx}

\begin{document}
\bibliographystyle{prsty}

\title{Quasiparticle Relaxation Across a Spin Gap in the Itinerant Antiferromagnet UNiGa$_{5}$}
\author{Elbert E. M. Chia}
\author{H. J. Lee}
\author{Namjung Hur}
\author{E. D. Bauer}
\author{T. Durakiewicz}
\author{R. D. Averitt}
\author{J. L. Sarrao}
\author{A. J. Taylor}
\affiliation{Materials Science and Technology Division, Los Alamos
National Laboratory, Los Alamos NM 87545, USA}
\date{\today}

\begin{abstract}
Ultrafast time-resolved photoinduced reflectivity is measured for
the itinerant antiferromagnet UNiGa$_{5}$ ($T_{N} \approx$85~K) from
room temperature to 10~K. The relaxation time $\tau$ shows a sharp
increase at $T_{N}$ consistent with the opening of a spin gap. In
addition, the temperature dependence of $\tau$ below $T_{N}$ is
consistent with the opening of a spin gap leading to a quasiparticle
recombination bottleneck as revealed by the Rothwarf-Taylor model.
This contrasts with canonical heavy fermions such as CeCoIn$_{5}$
where the recombination bottleneck arises from the hybridization
gap.
\end{abstract}

\maketitle

There has been a great deal of interest in the ``115'' series of
rare-earth and actinide compounds, such as CeTIn$_{5}$ (T=Co, Rh,
Ir) \cite{Hegger00,Petrovic2001b,Thompson01}, and PuT'Ga$_{5}$
(T'=Co, Rh) \cite{Sarrao02,Wastin03}, all of which exhibit
unconventional superconductivity. Attention has been drawn to the
uranium isomorphs UMGa$_{5}$ (M=Ni, Pd, Pt) which are isostructural
to the Ce and Pu counterparts, yet do not exhibit superconductivity
at either ambient or high pressure. It has been suggested that this
is due to the strong hybridization and itinerant character of the
5\textit{f} levels, which leads to a relatively wide 5\textit{f}
band at $E_{F}$ and a lack of spin fluctuations \cite{Kaneko03b}.
Understanding the evolution of behavior from the parent compound
UGa$_{3}$ to UMGa$_{5}$, especially across M from Ni to Pt with a
decreasing N$\acute{e}$el temperature, is illuminating for future
studies of delta-Pu and PuT'Ga$_{5}$. UMGa$_{5}$ also differs from
PuT'Ga$_{5}$ regarding the extent of itinerancy of the 5\textit{f}
electrons --- the former is almost fully itinerant, while the latter
is only partially itinerant. Our study thus sheds light on the
effect of the degree of itinerancy on the relaxation dynamics of
these materials.

Time-resolved photoinduced reflectivity measurements have been
performed on heavy fermions (HF) such as YbAgCu$_{4}$
\cite{Demsar03a} and heavy-fermion superconductors such as
CeCoIn$_{5}$ \cite{Demsar06b}. Both display a divergence of the
electron-phonon (e-ph) relaxation time $\tau$ at the lowest
temperatures, which can be explained by the Two-Temperature Model
(TTM) \cite{Groeneveld95}. This model describes the time evolution
of the electron ($T_{e}$) and lattice ($T_{l}$) temperatures by two
coupled differential equations. In this model (which assumes a
thermal electron distribution) $\tau$ varies as $T^{-1}$ at low
temperatures if there is no blocking of e-ph scattering of heavy
electrons within the density-of-states (DOS) peak, but varies more
strongly than $T^{-1}$, and has a larger magnitude, if there is e-ph
blocking within the DOS peak \cite{Ahn04}. On the other hand,
similar measurements have also been performed on materials with a
gap in the quasiparticle spectrum including high-temperature
superconductors like YBCO \cite{Han90,Demsar99c} and
charge-density-wave materials like K$_{0.3}$MoO$_{3}$
\cite{Demsar99b}. In these materials the relaxation time diverges
near the transition temperature when a gap opens in the
single-particle DOS. The opening of a gap can lead to a relaxation
bottleneck, arising in superconductors from the competition between
quasiparticle recombination and pair breaking by phonons
\cite{Demsar01}. However, at present there has been no measurements
of the relaxation dynamics of the antiferromagnetic (AF) phase,
where a spin gap opens up below the N$\acute{e}$el temperature
$T_{N}$.

In this Letter, we present time-resolved optical pump-probe data
where we measure photoinduced changes in the reflectivity ($\Delta
R/R$) of UNiGa$_{5}$ from room temperature down to 10~K. The decay
time of $\Delta R/R$ (directly related to electron-\textit{boson}
relaxation time $\tau$) increases sharply at $T_{N}$ and shows a
quasi-divergence below $T_{N}$, consistent with the opening of a
gap. Though $\tau \sim T^{-1}$ at low temperatures is suggestive of
HF behavior, our detailed TTM calculations show otherwise. Instead,
data over the entire temperature range of the AF phase could be fit
using the phenomenological Rothwarf-Taylor (RT) model
\cite{Rothwarf67}, where a boson bottleneck occurs due to the
opening of the \textit{spin gap}. We emphasize that the observed
dynamics arise from the opening of a spin gap and not a
hybridization gap as in HFs with larger mass renormalization.

UNiGa$_{5}$ is a 5\textit{f} itinerant antiferromagnet with
N$\acute{e}$el temperature $T_{N}$$\approx$85~K, and electronic
specific heat coefficient $\gamma$ = 30~mJ/mol.K$^{2}$
\cite{Tokiwa01}. The moderate value of $\gamma$ suggests that
UNiGa$_{5}$ is a marginal HF. Angle-resolved photoemission data
\cite{Durakiewicz06} of UNiGa$_{5}$ reveal a high intensity DOS peak
near $E_{F}$. Such peaks are characteristic of HFs. The
full-width-half-max (FWHM) of the feature is about 150~meV. A small
hump in the electrical resistivity at $T_{N}$ is reminiscent of the
spin density wave (SDW) formation \cite{Tokiwa01}, where a spin gap
forms.

Single crystals of UNiGa$_{5}$ were grown in Ga flux
\cite{Moreno05}, with dimensions $\sim$1mm x 1mm x 0.4mm. Specific
heat measurements were performed in a Quantum Design PPMS from 2~K
to 300~K. The photoinduced reflectivitiy measurements were performed
using a standard pump-probe technique \cite{Demsar99d}, with a
Ti:sapphire laser producing 100-fs pulses at approximately 800~nm
(1.5~eV) as the source of both pump and probe optical pulses. The
pump and probe pulses were cross-polarized. The experiments were
performed with a pump fluence of $\sim$1.0 $\mu$J/cm$^{2}$. The
probe intensity was approximately 25 times lower. Data were taken
from 10~K to 300~K. The photoinduced temperature rise at the lowest
temperatures was estimated to be $\sim$10~K (in all of the data the
$T$ increase of the illuminated spot has been accounted for).

The temperature dependence of the specific heat $C_{p}$ (not shown)
shows a small hump at $T_{N}$, indicative of the formation of a spin
gap. At low temperatures we fit the data to $C_{p}$ = $C_{e}$ +
$C_{l}$ = $\gamma T$ + $\beta T^{3}$, obtaining $\gamma$=21
mJ/mol.K$^{2}$ and $\beta$=0.34 mJ/mol.K$^{4}$. The values of
$C_{e}$ and $C_{l}$ will be used later in the TTM calculations.

\begin{figure} \centering \includegraphics[width=8cm,clip]{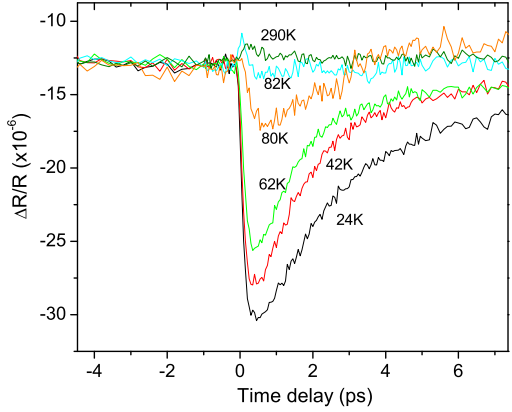}
\caption{Transient reflection $\Delta R/R$ from UNiGa$_{5}$ after
photoexcitation by a 100-fs laser pulse at a number of temperatures
above and below $T_{N}$.} \label{fig:AllT}
\end{figure}

In Figure~\ref{fig:AllT} we show the time dependence of the
photoinduced signal at a number of temperatures below and above
$T_{N}$. The time evolution of the photoinduced reflection $\Delta
R/R$ first shows a rapid rise time (of the order of the pump pulse
duration) followed by a subsequent picosecond decay. These data can
be fit using a single exponential decay over the entire temperature
range, $\Delta R/R = A \exp (-t/\tau)$.

\begin{figure} \centering \includegraphics[width=8cm,clip]{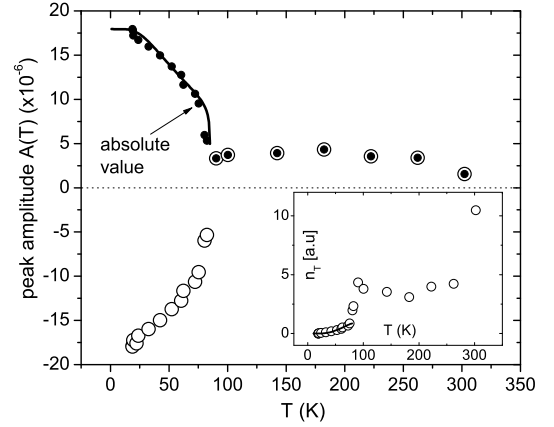}
\caption{(o): Amplitude of the transient component, $A(T)$. Solid
circles: $|A(T)|$. Solid line: Fit to $|A(T)|$ by
Eq.~\ref{eqn:amplitude} with $\Delta (0)$=0.55$k_{B}T_{N}$. Inset:
(o)=Temperature dependence of the QP density $n_{T}$. Solid line =
Fit to $n_{T}(T)$ by Eq.~\ref{eqn:nTSC} with $\Delta
(0)$=1.0$k_{B}T_{N}$.} \label{fig:amplitude}
\end{figure}

\begin{figure}
\centering \includegraphics[width=8cm,clip]{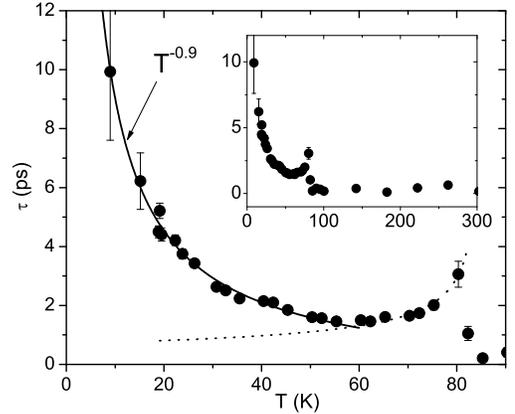}
\caption{Temperature dependence of relaxation time $\tau$. Solid
line: Fit of low-temperature data to $1/T^{n}$ dependence, yielding
$n$=0.9. Dotted line: Fit of data near $T_{N}$ to $\tau \propto
1/\Delta (T)$. Inset: $\tau$ over the entire temperature range.}
\label{fig:tau}
\end{figure}

The temperature dependence of the amplitude $A(T)$ and relaxation
time $\tau (T)$ of the photoinduced reflectivity are shown in
Figures~\ref{fig:amplitude} and \ref{fig:tau}, respectively. We
first analyze $A(T)$. We see that there is both a sign change and a
decrease in magnitude as we approach $T_{N}$ from below. The origin
of the sign change is under investigation. Its magnitude, however,
has been shown to be \cite{Demsar99b}
\begin{equation}
|A(T)| \propto \frac{\epsilon_{I}/(\Delta (T) + k_{B}T/2)}{1 + B
\sqrt{\frac{2 k_{B}T}{\pi \Delta (T)}}\exp (\frac{-\Delta
(T)}{k_{B}T})} \label{eqn:amplitude}
\end{equation} where $\Delta$ is the spin gap magnitude, $\epsilon_{I}$ the pump laser
intensity per unit cell, and $B$ is a constant. $\Delta$ is taken to
vary with temperature as $\Delta (T)=1.74 \Delta
(0)(1-T/T_{N})^{1/2}$, the mean-field result. Fitting the data in
Fig.~\ref{fig:amplitude} to Eq.~\ref{eqn:amplitude}, we obtain $B
\approx 5$ and $\Delta (0)$=47~K=0.55$k_{B}T_{N}$. This value for
$\Delta (0)$ agrees well with the value of 44~K from resistivity
data \cite{Moreno05}.

Turning to $\tau$ in Fig.~\ref{fig:tau}, we observe that $\tau$
diverges at the lowest temperatures and near $T_{N}$. The
low-temperature divergence follows a $T^{-1}$ dependence, initially
suggesting that the TTM applies with no e-ph blocking within the DOS
peak, in contrast to the presence of e-ph blocking in YbAgCu$_{4}$
\cite{Demsar03a,Ahn04}. Near $T_{N}$, $\tau$ is shown to vary with
temperature according to the expression \cite{Kabanov99}

\begin{equation}
\frac{1}{\tau_{ep}}=\frac{12 \Gamma_{\omega} \Delta (T)^{2}}{\hbar
\omega^{2} \ln \left[ \frac{1}{\frac{\epsilon_{I}}{2N(0)\Delta
(0)^{2}}+ \exp \left( \frac{-\Delta (T)}{k_{B}T} \right)} \right]}
\label{eqn:tau}
\end{equation}
where $\Gamma_{\omega}$ is the Raman phonon linewidth and $\omega$
the phonon energy. However, absent experimental values for $\omega$
and $\Gamma_{\omega}$, we use the expression $\tau \propto 1/\Delta
(T)$, which is valid near $T_{N}$ and is shown as a dotted line in
Fig.~\ref{fig:tau}.

\begin{figure}
\centering \includegraphics[width=8cm,clip]{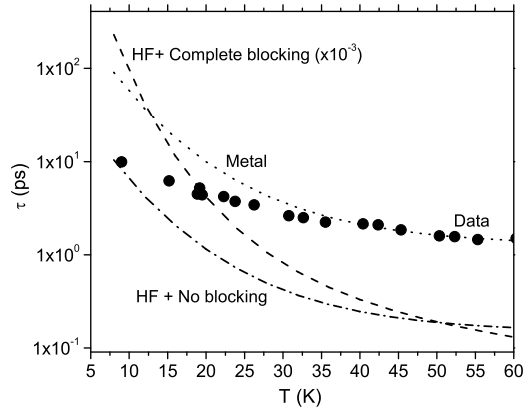}
\caption{Theoretical fits to relaxation time $\tau$. Solid circles:
data. Lines: Fit to Eqn.~\ref{eqn:TTMtau} for a metal (dotted), HF
with e-ph blocking (dashed), HF with no e-ph blocking
(dashed-dotted), inside the DOS peak.} \label{fig:tauallcases}
\end{figure}

In the TTM the relaxation time is given by \cite{Groeneveld95}
\begin{equation}
\frac{1}{\tau_{ep}}=g(T)\left( \frac{1}{C_{e}(T)} +
\frac{1}{C_{l}(T)} \right) \label{eqn:TTMtau}
\end{equation}
where $C_{e}$ and $C_{l}$ are the electronic and lattice specific
heat, respectively. $g(T)$ is termed the e-ph coupling function, and
is given by $g(T)=dG(T)/dT$, where
\begin{equation}
G(T)=4g_{\infty}\left(\frac{T}{\theta_{D}} \right)
\int^{\theta_{D}/T}_{0} dx \frac{x^{4}}{e^{x}-1} \chi (x,T)
\label{eqn:GT}
\end{equation}
\begin{equation}
\chi(x,T)=\frac{1}{xT} \int^{\infty}_{-\infty} d\epsilon
\frac{D_{e}(\epsilon)D_{e}(\epsilon^{\prime})F(\epsilon,\epsilon^{\prime})}{D_{0}^{2}}
\{f_{0}(\epsilon)-f_{0}(\epsilon^{\prime})\}. \label{eqn:chiT}
\end{equation}Here $g_{\infty}$ is the e-ph coupling constant, $\theta_{D}$ is the
Debye temperature ($\theta_{D}$=260~K for UNiGa$_{5}$ from
specific-heat data), $\epsilon^{\prime}=\epsilon + \xi$, and $\xi =
xT$. $\xi (x,T)$ is included to account for the variation in the
electronic DOS $D_{e}(\epsilon)$ and the normalized e-ph scattering
strength $F(\epsilon , \epsilon^{\prime})$ over the energy range
$E_{F} \pm \hbar \omega_{D}$, where the detailed procedure is
described in Ref.~\onlinecite{Demsar03a}. In normal metals such as
Au or Ag, $D_{e}^{metal}(\epsilon)=D_{0}$, $F$=1, giving $\xi$=1.
For UNiGa$_{5}$, the phonon [$D_{p}(\omega)$] and electron
[$D_{e}(\epsilon)$] DOS were chosen such that they fit the specific
heat and photoemission data. For the electron DOS, we used
$D_{e}^{HF}(\epsilon) = D_{peak} \exp[-(\epsilon/\Delta)^{2}] +
D_{0}$, where $D_{peak}$=3.9 eV$^{-1}$f.u.$^{-1}$ spin$^{-1}$,
$\Delta$=0.18 eV, and $D_{0}$=2.0 eV$^{-1}$f.u.$^{-1}$ spin$^{-1}$.
For the phonon DOS, we used the Debye model $D_{p}(\omega) \sim
\omega^{2}$ with $\omega_{D}$=260~K. The lines of
Figure~\ref{fig:tauallcases} show the calculated values of $\tau$
using Equations 2-4 for 3 cases: (a) metal (dotted line), (b) HF
with e-ph blocking within the DOS peak (dashed line), (c) HF with no
e-ph blocking within the DOS peak (dashed-dotted line). It is clear
that they do not agree with the experimental data, especially at the
lowest temperatures where the slopes do not match, regardless of the
value of $g_{\infty}$, showing that the both the metal and HF
picture (with or without blocking) cannot account for the
temperature dependence of $\tau$. In addition, recent investigations
have called into question the validity of the TTM when a gap opens
up in the DOS \cite{Demsar06a,Demsar06b}.

We turn next to the Rothwarf-Taylor (RT) model \cite{Rothwarf67}.
This phenomenological model was used to describe the relaxation of
photoexcited superconductors \cite{Demsar03c}, where the presence of
a gap in the electronic DOS gives rise to a relaxation bottleneck
for carrier relaxation. When two quasiparticles (QP) with energies
$\geq \Delta$, where $\Delta$ is the gap magnitude, recombine, a
high-frequency phonon (HFP) ($\omega > 2\Delta$) is created. Since a
HFP can subsequently break a Cooper pair creating two QPs, the SC
recovery is governed by the decay of the HFP population. The
evolution of  QP and HFP populations are described by a set of two
coupled nonlinear-differential equations:
\begin{eqnarray}
dn/dt & = & \eta N - R n^{2} \\
dN/dt & = & -\eta N/2 + R n^{2}/2 - \gamma (N-N_{T}),
\label{eqn:RT2}
\end{eqnarray}

Here $n$ and $N$ are the concentrations of QP and HFPs,
respectively, $\eta$ is the probability for QP creation by HFP
absorption, and $R$ the rate of QP recombination with the creation
of a HFP. $N_{T}$ is the concentration of HFPs in thermal
equilibrium, and $\gamma$ their decay rate (governed either by
anharmonic decay or by diffusion out of the excitation volume). The
RT model has also been applied to the study of HFs, where the
dynamics are associated with a gap resulting from the hybridization
of the conduction electrons with the localized \textit{f}-levels
\cite{Demsar06a}. In the following, we apply the RT model to
UNiGa$_{5}$.

\begin{figure}
\centering \includegraphics[width=8cm,clip]{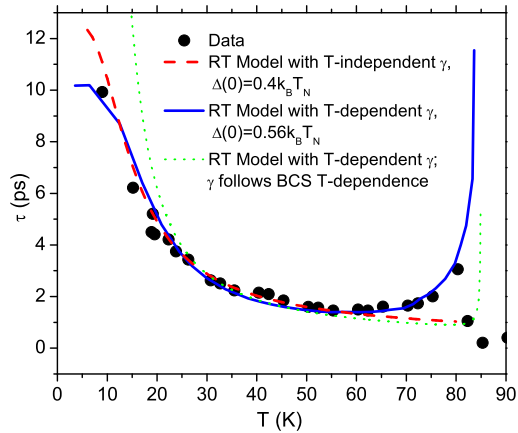}
\caption{Solid circles: Temperature dependence of relaxation time
$\tau$. Dashed line: Fit to Eq.~\ref{eqn:RTtau} assuming a
$T$-independent $\Gamma$. Solid line: Fit to Eq.~\ref{eqn:RTtau}
assuming a $T$-dependent $\Gamma$ deduced from neutron scattering
data. Dotted line: Fit to Eq.~\ref{eqn:RTtau} assuming a
$T$-dependent $\Gamma$ that follows a BCS dependence.}
\label{fig:UNiGa5tauRTfit5}
\end{figure}

The results of the RT model are as follows
\cite{Kabanov05,Demsar06b}: from the $T$-dependence of the amplitude
$A$, one obtains the density of thermally excited QPs $n_{T}$ via
the relation
\begin{equation}
n_{T}(T) \propto  \mathcal{A}^{-1} -1 \label{eqn:nTA}
\end{equation} where $\mathcal{A}(T)$ is the normalized amplitude
(normalized to its low temperature value, $\mathcal{A}(T)=A(T)/A(T
\rightarrow 0)$. Then, from the QP density per unit cell at
temperature $T$ \cite{Kabanov99}
\begin{equation}
n_{T}(T) \propto  \sqrt{\Delta (T) T} \exp (-\Delta (T)/T)
\label{eqn:nTSC}
\end{equation} one can extract the value of the energy gap.
The inset of Fig.~\ref{fig:amplitude} shows the $T$-dependence of
the peak amplitude $A(T)$, and the QP density $n_{T}$ calculated
from Eq.~\ref{eqn:nTA}. A fit of $n_{T}(T)$ at low temperatures to
Eq.~\ref{eqn:nTSC} yields $\Delta (0)=1.0k_{B}T_{N}$. Moreover, for
a constant pump intensity, the $T$-dependence of $n_{T}$ also
governs the $T$-dependence of $\tau^{-1}$, given by
\begin{equation}
\tau^{-1}(T) =  \Gamma [\delta (\beta n_{T} +1)^{-1} + 2 n_{T}]
\label{eqn:RTtau}
\end{equation} where $\Gamma$, $\delta$ and $\beta$ are $T$-\textit{independent} fit
parameters. However, from Ref.~\onlinecite{Kabanov05}, we see that
this expression for $\tau$ fit the data only up to $\sim$0.8$T_{c}$
and fails to reproduce the upturn in $\tau$ near $T_{c}$. The
$T$-dependence of $\tau$ near $T_{c}$ can be explained by taking
into account the $T$-dependence of $\gamma$, the phonon decay rate
\cite{Kabanov99,Kabanov05}. We include this $T$-dependence of
$\gamma$ into $\Gamma (T)$, where $\Gamma(T) \propto \Delta (T)$
\cite{Kabanov99}, the (spin) gap magnitude, whose $T$-dependence can
be obtained from neutron scattering data \cite{Tokiwa02c}.
Fig.~\ref{fig:UNiGa5tauRTfit5} shows the fit of Eq.~\ref{eqn:RTtau}
to data assuming a $T$-independent $\Gamma$ (dashed line) and a
$T$-dependent $\Gamma$ (solid line). It is clear that using a
$T$-dependent $\Gamma$ in Eq.~\ref{eqn:RTtau} matches data from
low-$T$ to $T_{N}$, for $\Delta (0)=0.6k_{B}T_{N}$, in agreement
with resistivity data. It is important to note that this fit uses
the $T$-dependence of the \textit{spin} gap from neutron scattering
data, which differs from the BCS $T$-dependence from which a good
fit is not possible (dotted line in Fig.~\ref{fig:UNiGa5tauRTfit5}).
This strongly suggests that our data for $\tau$ arises from some
sort of \textit{boson} (most likely phonon) bottleneck due to the
presence of the spin gap, as one enters the AF phase. Since our
pump-probe setup only enables the observation of a gap in the charge
channel, not the spin channel, the fact that we were able to observe
the feature at $T_{N}$ in UNiGa$_{5}$ suggests that a
\textit{spin-driven charge gap}, whose temperature dependence of its
magnitude follows that of the spin gap, also opens up at $T_{N}$.
Since the TTM does not account for a gap opening at the Fermi level,
the presence of a gap below $T_{N}$ in UNiGa$_{5}$ probably explains
why the TTM fails to explain our data. One could argue that what we
are seeing here is the hybridization gap. However, the fact that (a)
we see a gap open up at $T_{N}$, and (b) we have used the
temperature dependence of the spin gap (deduced from neutron
scattering data) in our fit to the RT model imply that it is
probably a spin-driven charge gap that opens up at $T_{N}$. It seems
fortuitous that a hybridization gap would open up at $T_{N}$.
Moreover, the (almost) fully itinerant nature of UNiGa$_{5}$ implies
a very small hybridization gap, consistent with the small value of
$\gamma$, suggesting that one can observe the hybridization gap only
at very low temperatures outside the temperature range of our setup.

We have performed time-resolved photoinduced reflectivity
measurements on the itinerant antiferromagnet UNiGa$_{5}$ ($T_{N}
\approx$85~K) from room temperature to 10~K. The relaxation time
$\tau$ increases sharply near $T_{N}$, which we attribute to the
opening of a spin gap. In addition, we fit the data over the entire
temperature range of the AF phase using the Rothwarf-Taylor model
\cite{Rothwarf67}, where a boson bottleneck occurs due to the
opening of the spin gap, rather than a hybridization gap as found
for heavy fermions. The transient amplitude exhibits a sign change
at $T_{N}$, whose temperature dependence is also consistent with the
appearance of a spin gap.

E.E.M.C. appreciates useful conversations with J. Demsar, D.
Mihailovic, J. Thompson, S. Trugman and M. Salamon. Work at Los
Alamos National Laboratory was supported by the U.S. Department of
Energy and the Los Alamos LDRD program. E.E.M.C. acknowledges
postdoctoral fellowship support from the G. T. Seaborg Institute for
Transactinium Science.

\bibliographystyle{prsty}
\bibliography{UMGa5,Ultrafast,PMGa5,CeCoIn5v11,RNBC}
\bigskip

\end{document}